# Optical helicity of unpolarized light

Kayn A. Forbes

*School of Chemistry, University of East Anglia, Norwich, NR20 7TJ, U.K.*

k.forbes@uea.ac.uk

Recently Eismann, et al. 2021 *Nat. Photonics* **15** 156–161 showed that the extraordinary transverse spin momentum density of spatially confined optical fields is largely independent of polarization. Here it is shown that 3D structured optical vortices which possess the phase factor $\exp(i\ell\phi)$ have a contribution to the optical helicity density which is completely independent of polarization. In stark contrast to what is known in classical optics with plane waves and paraxial light, the physical consequence is that unpolarized light can exhibit optical activity and chiral light-matter interactions.

Non-paraxial optical fields exhibit fascinating properties compared to the plane waves and paraxial fields which have dominated classical optics and light-matter interactions for decades. In recent years, with the growth of modern nano-optics and photonics, the extraordinary properties of non-paraxial fields have found widespread utilization [1,2]. In this Letter we refer to paraxial modes and propagating plane waves as 2D structured fields, in that they may possess inhomogeneous spatial or polarization degrees of freedom in the transverse (*x*,*y*) plane, but are homogenous in the direction of propagation (*z*) [3]. 3D structured optical fields are inhomogeneous along their direction of propagation, examples include evanescent waves or tightly focussed laser beams, the essential requirement being electromagnetic fields that are spatially confined in some way or another. A crucial difference between 2D and 3D structured light is that the latter possess significant longitudinal components with respect to the direction of propagation of their electromagnetic fields in addition to the usually dominant transverse fields. The non-paraxial nature and 3D structure of spatially confined optical fields has led to some remarkable light-matter interactions [4–7], which are especially striking when compared to the prevailing textbook description of light-matter interactions in terms of propagating plane waves.

One of the extraordinary properties of non-paraxial optical fields is that they possess a transverse spin momentum, orthogonal to the main direction of propagation [8]. Eismann, et al. [9] have recently proven both theoretically and experimentally that this transverse spin momentum is largely independent of polarization, and remarkably may be generated even with unpolarized optical fields. Closely related to spin angular momentum is the optical helicity, and in fact both are related to one another by a continuity equation [10–13]. In the plane wave case, it is well known that optical helicity is proportional to the degree of circular polarization: it is zero for linearly or randomly polarized optical fields and takes on its maximum value for circular polarization. The optical helicity is proportional to the optical chirality for monochromatic fields and is responsible for optical activity and chiral light-matter interactions [14–21]. Here we show that 3D structured light beams generated from a 2D source which possesses the azimuthal phase factor $\exp(i\ell\phi)$, commonly referred to as optical vortices or twisted light [22], acquire a non-zero optical helicity density that is fully independent of the 2D polarization, occurring even for unpolarized light. The striking physical consequence of this is that unpolarized light can exhibit optical activity.

Optical helicity $\mathcal{H}$ in the Coulomb gauge is defined as [10,12,18]

$$\mathcal{H} = \int d^3r \frac{\varepsilon_0 c}{2}\left(A^\perp \cdot B - C^\perp \cdot E\right), \quad (1)$$

where $A^\perp$ $C^\perp$ are the vector potentials and $E$ $B$ the electric and magnetic field, respectively. The total helicity $\mathcal{H}$ (1) is both gauge and Lorentz invariant, however the integrand, which represents the optical helicity density $h$, is not Lorentz invariant. Lorentz invariance is sacrificed to make $h$ gauge invariant, important for calculating experimentally determinable physical quantities in optics [13]. The cycle-averaged helicity density $\bar{h}$ for classical fields is

$$\bar{h} = -\frac{\varepsilon_0 c}{2\omega}\text{Im}\left(E^* \cdot B\right). \quad (2)$$

Clearly both expressions (1)-(2) are conserved quantities of the field, i.e. $\dot{\mathcal{H}}, \dot{h}, \dot{\bar{h}} = 0$. For 2D structured light, propagating plane waves, and even evanescent waves the optical helicity is directly proportional to the degree of circular polarization $\bar{h} \propto \sigma$, where $\sigma = \pm 1$ for circularly polarised light and is zero for both linearly and unpolarized light [4].

In this work we are mainly interested in calculating $\bar{h}$ for unpolarized 3D structured Laguerre-Gaussian (LG) modes. The simplest way to calculate this is to average the conserved quantity over two orthogonal polarization states on the Poincaré sphere. We therefore choose to average over two electric fields, one linearly polarized in the $x$ direction and the other in the $y$ direction (with the corresponding magnetic fields polarized in the $y$ and $-x$ directions, respectively).

The electric field for an $x$ polarized monochromatic 3D LG mode in the first post-paraxial approximation is [23,24]

$$\boldsymbol{E}_{\text{LG}} = E_0 \left\{ \hat{\boldsymbol{x}} + \hat{\boldsymbol{z}} \frac{i}{k} \left( \gamma \cos\phi - \frac{i\ell}{r} \sin\phi \right) \right\} u_{\ell,p}^{\text{LG}}(r,\phi,z), \tag{3}$$

where $E_0$ is the field amplitude, $\gamma = \left\{ \frac{|\ell|}{r} - \frac{2r}{w^2(z)} + \frac{ikr}{R(z)} - \frac{4r}{w^2(z)} \frac{L_{p-1}^{|\ell|+1}}{L_p^{|\ell|}} \right\}$, and $u_{\ell,p}^{\text{LG}}(r,\phi,z)$ is [25]

$$u_{\ell,p}^{\text{LG}}(r,\phi,z) \sqrt{\frac{2p!}{\pi w_0^2 (p+|\ell|!)}} \frac{w_0}{w(z)} \left( \frac{\sqrt{2}r}{w(z)} \right)^{|\ell|} L_p^{|\ell|}\left[ \frac{2r^2}{w^2(z)} \right] \exp(-r^2/w^2(z))$$
$$\exp i\left( kz + \ell\phi - \omega t + kr^2/2R(z) - (2p+|\ell|+1)\xi(z) \right), \tag{4}$$

$\ell \in \mathbb{Z}$ is the pseudoscalar topological charge; $p$ is the radial index, all other quantities have their usual meanings. The $\hat{\boldsymbol{x}}$ dependent term in (3) is the transverse field component and taken alone represents a 2D structured LG mode; the $\hat{\boldsymbol{z}}$ dependent term is the longitudinal field, responsible for 3D structure. The corresponding magnetic field is given by

$$\boldsymbol{B}_{\text{LG}} = B_0 \left\{ \hat{\boldsymbol{y}} + \hat{\boldsymbol{z}} \frac{i}{k} \left( \gamma \sin\phi + \frac{i\ell}{r} \cos\phi \right) \right\} u_{\ell,p}^{\text{LG}}(r,\phi,z). \tag{5}$$

Inserting (3) and (5) into (2) gives the optical helicity density as

$$\bar{h}^x = -\text{Re}\, \frac{I(r,z)}{c\omega} \frac{\ell}{k^2 r} \gamma, \tag{6}$$

where $I(r,z) = \frac{c\varepsilon_0}{2} E_0^2 |u_{\ell,p}^{\text{LG}}|^2$ is the input beam intensity. It is crucial to realize that this optical helicity density stems purely from the longitudinal field components and is proportional to $\ell$, and so is unique to 3D structured beams that possess an azimuthal phase $\exp(i\ell\phi)$ and is not exhibited by 2D structured light. Optical helicity of a 2D structured field comes purely from the degree of circular polarization of the transverse fields.

For the orthogonally polarized beam the electric and magnetic fields are

$$\boldsymbol{E}_{\text{LG}} = E_0 \left\{ \hat{\boldsymbol{y}} + \hat{\boldsymbol{z}} \frac{i}{k} \left( \gamma \sin\phi + \frac{i\ell}{r} \cos\phi \right) \right\} u_{\ell,p}^{\text{LG}}(r,\phi,z), \tag{7}$$

$$\boldsymbol{B}_{\text{LG}} = B_0 \left\{ -\hat{\boldsymbol{x}} - \hat{\boldsymbol{z}} \frac{i}{k} \left( \gamma \cos\phi - \frac{i\ell}{r} \sin\phi \right) \right\} u_{\ell,p}^{\text{LG}}(r,\phi,z). \tag{8}$$

Inserting (7) and (8) into (2) gives

$$\bar{h}^y = -\text{Re}\, \frac{I(r,z)}{c\omega} \frac{\ell}{k^2 r} \gamma, \tag{9}$$

and for unpolarized $(n)$ light the optical helicity density $\bar{h}^n$ is therefore

$$\bar{h}^n = \frac{\bar{h}^x + \bar{h}^y}{2} = \bar{h}^x. \tag{10}$$

This optical helicity density is fully independent of the polarization of the transverse field and is plotted in Figure 1. For example, taking a paraxial 2D structured vortex and tightly focusing it to create a 3D structure, the ensuing optical helicity density generated is completely independent of the 2D polarization state of the input paraxial structured mode. It is important to emphasize that this optical helicity density is proportional to $\ell$ and so linearly/randomly polarized Gaussian or Hermite-Gaussian beams, for example, do not possess this optical helicity regardless of whether they are 2D or 3D structured. The input 2D beam must possess the phase factor $\exp(i\ell\phi)$, so Bessel beams, for example, would also display this polarization independent optical helicity density. The reason for this is that the origin of the longitudinal fields of 3D modes which generate this polarization independent optical helicity density are in the gradients of the transverse fields, i.e. $\propto \partial_r + \partial_\phi$, and it is the azimuthal gradient which provides the unique $\ell$-dependent longitudinal component.

Our result can be derived via an alternative method. The optical helicity density generated for a 3D field generated from a circularly polarized 2D structured LG mode is [24]

$$\bar{h}^\sigma = \frac{I(r,z)}{c\omega} \left( \sigma + \frac{1}{2k^2} \left[ \sigma (\text{Re}\,\gamma)^2 - \frac{2\ell}{r} \text{Re}\,\gamma + \frac{\ell^2 \sigma}{r^2} \right] \right), \tag{11}$$

where $\sigma = \pm 1$, the positive sign corresponding to a left-handed input and the negative sign right-handed. Adding the orthogonal polarizations $\sigma = 1$ and $\sigma = -1$ together and averaging the result gives the unpolarized result:

$$\bar{h}^n = \frac{\bar{h}^{|\sigma|} + \bar{h}^{-|\sigma|}}{2} = -\text{Re}\, \frac{I(r,z)}{c\omega} \frac{\ell}{k^2 r} \gamma. \tag{12}$$

This gives the identical optical helicity density to the method of averaging orthogonal linear polarizations. We can conclude that there is a contribution to optical helicity density which is purely polarization independent in 3D vortex beams, even when the input beam has a degree of circular polarization in the transverse fields and it is therefore robust again spin-orbit-interactions of light [5]. It is important to note that this optical helicity density we are concerned with does not contribute to the integrated value of optical helicity:

$$\mathcal{H} = \int \bar{h} d^3 \mathbf{r} = 0. \tag{13}$$

Its experimental observation therefore requires particles smaller than the transverse dimension of the focal field.

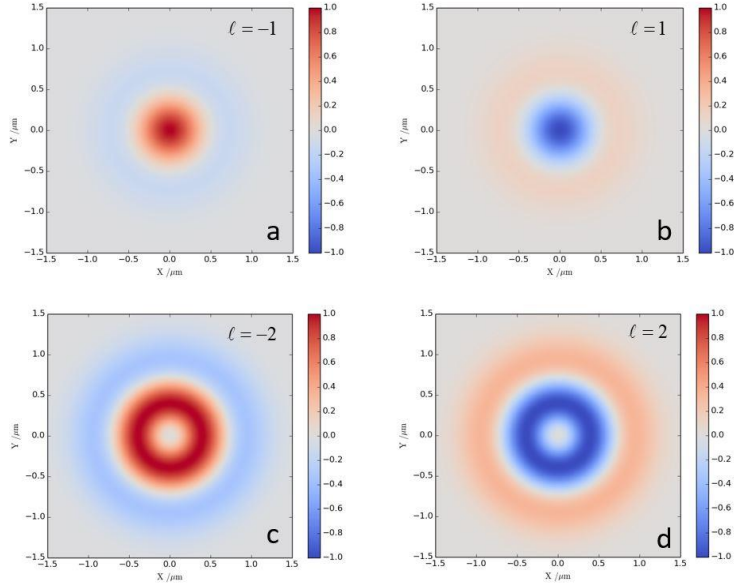

**Figure 1**: normalized polarization-independent optical helicity density (10) at $w_0(z=0)$. $p=0$ in (a)-(d)

As mentioned above, recently it was theoretically and experimentally shown that transverse spin angular momentum is largely independent of the polarization state of the input 2D field in both focussed light and evanescent waves [9]. In their work the authors make the comment that 'However, in our case of an unpolarized source, the helicity and longitudinal spin vanish.' We are able to directly compare our results here to those in [9]. The fundamental Gaussian mode is simply $LG_{00}$ and $\ell=0$ in (12) shows that for a Gaussian beam $\bar{h}=0$ in agreement with [9]. The input 2D field which generates the 3D Gaussian field and evanescent wave in [9] responsible for the polarization independent transverse spin and zero helicity density does not possess the phase factor $\exp(i\ell\phi)$. An alternative physical interpretation to that given below

(10) is that this azimuthal phase leads to the canonical momentum density having an azimuthal component, which when projected onto the transverse spin angular momentum density leads to a non-zero helicity density for unpolarized light as we have shown here. In the case of a Gaussian beam [4] or evanescent wave [26] the canonical momentum density is purely in the longitudinal direction, and so even though these optical fields may possess a transverse spin density, its projection on their canonical momentum density to yield the helicity gives zero in any circumstances.

Another point worth comparing between optical helicity density and transverse spin momentum density is that in [9] it is stated that the transverse spin angular momentum is 'largely independent of the polarization state'. The dual symmetric transverse spin $s_\perp = s_\perp^E + s_\perp^B$ is fully independent of the polarization state, but the corresponding electric and magnetic spatial distributions do depend on the input 2D polarization, and this has important consequences in experimental observations due to the electric-bias nature of most materials (see *supplementary information*). However, the optical helicity density studied in this article is fully independent of polarization in every respect, including its interaction with matter. The electric-magnetic asymmetry of matter does not influence optical helicity because by its very nature it couples to the interferences between electric and magnetic dipoles of chiral matter. This agrees with the fact that the dual symmetric and dual asymmetric optical helicities are identical [12,24].

In this Letter we have highlighted a fully polarization independent optical helicity density, which exists even for unpolarized light, and that its generation requires the incident optical field to possess the phase factor $\exp(i\ell\phi)$. The input 2D fields with this phase structure before spatial confinement possess zero optical helicity but they are geometrically chiral, i.e. they possess a non-zero Kelvin's chirality [27]. In contrast, the Kelvin's chirality of a 2D Gaussian beam or evanescent field is zero. As such, it is only optical fields which possess a non-zero Kelvin's chirality which have the capacity to generate a polarization independent optical helicity density. Interestingly Kelvin's chirality does not interact (in a non-mechanical, spectroscopic sense) with chiral matter in the dipole approximation; multipole couplings of electric quadrupole nature or higher are required [28]. However, the polarization independent optical helicity density generated by these beams with Kelvin's chirality does interact via the interferences of electric and magnetic dipoles [29–31]. Is it therefore legitimate to suggest that there is a relationship between Kelvin's chirality and optical helicity, and in certain scenarios the two become coupled? This is what Nechayev et al. [27] propose. The problem is that just like geometrical chirality of matter, there is seemingly no satisfactory way to quantify the scale-dependent Kelvin's chirality, and fundamentally there is of course no quantum operator for chirality.

Experimental observation of the polarization independent contribution to the optical helicity density would be straightforward. For example, the experimental set up in Ref. [30] could very easily be modified to prove that 3D optical vortices possess a polarization independent optical helicity by simply comparing the transmission of unpolarized LG modes of $\ell = \pm 1, p = 0$ to that of any polarized Gaussian mode. It is known that a number of light-matter interactions are proportional to the optical helicity density [32], such as chiral optical forces, Rayleigh and Raman optical activity, and vortex dichroism [33], all of which provide a means to observe optical activity with unpolarized light, something which could never be envisaged with plane wave or paraxial light sources. This work therefore adds to the rapidly growing field of structured light and chirality [29].

K A F is grateful to the Leverhulme Trust for funding him through a Leverhulme Trust Early Career Fellowship (Grant Number ECF-2019-398) and thanks David L. Andrews for helpful discussions.